\documentclass[aps,prl,reprint,superscriptaddress
]{revtex4-1}
\usepackage{graphicx}
\usepackage{dcolumn}
\usepackage{rotating}
\usepackage{amssymb}
\usepackage{mathptmx}
\usepackage{amsfonts}
\usepackage{amsmath}
\usepackage{bm}
\usepackage{gensymb}

\begin{document}
\title{Neutron diffraction of field-induced magnon condensation in the spin-dimerized antiferromagnet Sr$_{3}$Cr$_{2}$O$_{8}$}

\author{Alsu Gazizulina}
\email{alsu.gazizulina@helmholtz-berlin.de}
\affiliation{Helmholtz Zentrum Berlin f\"{u}r Materialien und Energie, 14109 Berlin, Germany}
\affiliation{Physik-Institut, Universit\"{a}t Z\"{u}rich, Winterthurerstrasse 190, 8057 Z\"{u}rich, Switzerland}

\author{Diana Lucia Quintero-Castro}
\email{diana.l.quintero@uis.no}
\affiliation{Helmholtz Zentrum Berlin f\"{u}r Materialien und Energie, 14109 Berlin, Germany}
\affiliation{Department of Mathematics and Physics, University of Stavanger, 4036 Stavanger, Norway}

\author{Zhe Wang}
\affiliation{Fakult\"{a}t Physik, Technische Universit\"{a}t Dortmund, 44227 Dortmund, Germany}
\affiliation{Institute of Radiation Physics, Helmholtz-Zentrum Dresden-Rossendorf, 01328 Dresden, Germany}

\author{Fabienne~Duc}
\affiliation{Laboratoire National des Champs Magn\'etiques Intenses, CNRS-INSA-UGA-UPS, 31400 Toulouse, France}

\author{Frederic~Bourdarot}
\affiliation{Universit\'e Grenoble Alpes, CEA, IRIG, MEM, MDN, 38000 Grenoble, France}

\author{Karel Prokes}
\affiliation{Helmholtz Zentrum Berlin f\"{u}r Materialien und Energie, 14109 Berlin, Germany}

\author{Wolfgang Schmidt}
\affiliation{Forschungszentrum J\"{u}lich GmbH, J\"{u}lich Centre for Neutron Science at ILL, 38042 Grenoble, France}

\author{Ramzy Daou}
\affiliation{Laboratoire de Cristallographie et Sciences des Mat\'eriaux (CRISMAT), Normandie Universit\'e, UMR6508 CNRS, ENSICAEN, UNICAEN, 14000, Caen, France}
\affiliation{Max Planck Institute for the Chemical Physics of Solids, 01187 Dresden, Germany}

\author{Sergei Zherlitsyn}
\affiliation{Hochfeld-Magnetlabor Dresden (HLD-EMFL), Helmholtz-Zentrum Dresden-Rossendorf, 01328 Dresden, Germany}

\author{Nazmul Islam}
\affiliation{Helmholtz Zentrum Berlin f\"{u}r Materialien und Energie, 14109 Berlin, Germany}

\author{Nils Henrik Kolnes}
\affiliation{Department of Mathematics and Physics, University of Stavanger, 4036 Stavanger, Norway}

\author{Abhijit Bhat Kademane}
\affiliation{Department of Mathematics and Physics, University of Stavanger, 4036 Stavanger, Norway}

\author{Andreas Schilling}
\affiliation{Physik-Institut, Universit\"{a}t Z\"{u}rich, Winterthurerstrasse 190, 8057 Z\"{u}rich, Switzerland}

\author{Bella Lake}
\affiliation{Helmholtz Zentrum Berlin f\"{u}r Materialien und Energie, 14109 Berlin, Germany}
\affiliation{Institut f\"{u}r Festk\"{o}rperphysik, Technische Universit\"{a}t Berlin, 10623 Berlin, Germany}

\begin{abstract}
In this work, we investigate the evolution and settling of magnon condensation in the spin-1/2 dimer system Sr$_{3}$Cr$_{2}$O$_{8}$ using a combination of magnetostriction in pulsed fields and inelastic neutron scattering in a continuous magnetic field. The magnetic structure in the Bose-Einstein condensation (BEC) phase was probed by neutron diffraction in pulsed magnetic fields up to 39~T. The magnetic structure in this phase was confirmed to be an XY-antiferromagnetic structure validated by irreducible representational analysis. The magnetic phase diagram as a function of an applied magnetic field for this system is presented. Furthermore, zero-field neutron diffraction results indicate that dimerization plays an important role in stabilizing the low-temperature crystal structure.
\end{abstract}
\maketitle	

\section{\label{Intr}Introduction}

Novel states of matter associated with a quantum phase transition can be induced via tuning an external parameter.
In an increasing number of spin-dimerized quantum antiferromagnets, magnonic BEC has been observed at a magnetic field-induced quantum phase transition~\cite{zapf}.
However, as shown by the iconic representation of the spin-dimer systems in Ref.~\cite{zapf}, most of the observed magnonic BEC occurs in a relatively high magnetic field ($>10$~T), which can be even above 30~T, e.g. for Sr$_{3}$Cr$_{2}$O$_{8}$.
This places challenges on an experimental study of the microscopic properties of the exotic BEC phases.
In particular, a direct probe of the magnetic ordering structure using a diffraction technique is not a straightforward task.
Recent advances on sample environment for neutron scattering have been opening research in new areas of physics~\cite{Grenier}. Especially, the cryogenic conditions with a pulsed-field of 40~T operated at the Institute Laue Langevin~\cite{40T_art} allow us to directly access the field-induced magnonic BEC phase in Sr$_{3}$Cr$_{2}$O$_{8}$ via neutron scattering.

Sr$_{3}$Cr$_{2}$O$_{8}$ is a spin-1/2 dimerized antiferromagnet. The spin dimers consisting of $S$-1/2 Cr$^{5+}$ ions are coupled antiferromagnetically by an intra-bilayer exchange interaction $J_{0}$ along $c$-axis~\cite{chapon, singh}.
At room temperature, the single $3d^{1}$ electron of the Cr$^{5+}$ ion has a two-fold orbital degeneracy, which makes the system Jahn-Teller active. The Jahn-Teller transition is related to a structural phase transformation from hexagonal $R\bar{3}m$ to monoclinic $C2/c$ at $T_{\text{JT}} = 285$~K~\cite{chapon, wang_orb, dirk}. 
The magnetic frustration due to the hexagonal structure is lifted because of this distortion, leading to spatially anisotropic magnetic interactions~\cite{diana_mag}. 
The magnetic phase diagram of this compound is characterized by a lower critical field for the magnon condensation at B$_{c1}=$ 30.9(4)~T and an upper critical field at B$_{c2}=$ 61.9(3)~T~\cite{aczel_sr}. 
The maximum temperature of this dome-like phase is $T_{c} \approx$~8~K, at approximately 45~T. 
Recently, it was shown that at temperatures higher than $T_{c}$ (up to a maximum of 18~K), a strongly correlated magnonic liquid regime appears~\cite{wang_add}. 
The extracted critical exponent of the ordering temperature as a function of reduced field around B$_{c1}$, $T_{c} \sim$ (B - B$_{c1})^{\nu}$, was found to be $\nu =0.65$, very close to $\nu = 2/3\approx 0.67$ predicted for a three-dimensional BEC of magnons~\cite{aczel_sr}. This result is in contrast to the inconclusive results for the isostructural compound Ba$_{3}$Cr$_{2}$O$_{8}$ where $\nu$ was found to lie between 0.5 and $0.67$~\cite{aczel_ba, kofu}. These differences might be due to the weak Dzyaloshinskii–Moriya interactions in Sr$_{3}$Cr$_{2}$O$_{8}$ in comparison to those in Ba$_{3}$Cr$_{2}$O$_{8}$ \cite{wang_esr}. Recent experiments of dilatometry and ultrasound have revealed a strong spin-lattice coupling with the shrinkage of the unit cell in the ordered phase~\cite{nomura2020}.

In this work, we present a study of the field-induced magnonic BEC phase in Sr$_{3}$Cr$_{2}$O$_{8}$, by performing neutron diffraction in pulsed magnetic fields up to 40~T and inelastic neutron scattering in continuous magnetic fields. We have not only obtained the evolution of the spin excitations as a function of the applied magnetic field but more importantly, also been able to resolve neutron diffraction of the magnon condensation above B$_{c1}$. Thereby we provide further evidence for the field-induced quantum phase transition and determine the possible ordered magnetic structures in the BEC phase.

\section{\label{sec:one}Experimental details}

Single crystals of Sr$_{3}$Cr$_{2}$O$_{8}$ were grown using floating zone technique~\cite{growth_Sr_nazmul}. The atomic structure of this spin dimer system has been studied by performing single-crystal neutron diffraction on a 2-axis-diffractometer (E4) with an incident wavelength of 0.244~nm at Helmholtz Zentrum Berlin. 
In order to investigate the splitting of the magnonic states as a function of an applied magnetic field, INS experiments were performed at 2~K at a cold neutron three-axis spectrometer IN12 at the Institute Laue-Langevin in Grenoble~\cite{in12}. Constant wave vector energy scans were carried out at zero magnetic fields and 15~T. The magnetic field was applied perpendicular to the (h, h, l)$_{h}$ scattering plane. For these measurements, the final wavevector was fixed to 1.5~\AA$^{-1}$. 

The high field magnetostriction data have been taken at the Hochfeld-Magnetlabor Dresden. Optical fibre strain gauges based on fibre Bragg gratings were used to measure the strain of the sample~\cite{ramzy}. The measurements were done using a pulsed magnetic field up to 64~T applied along the $a$-axis and a $^3$He insert was used to reach 0.69~K.

The high field neutron diffraction experiment was performed on the thermal neutron three-axis spectrometer IN22, a Collaborating Research Group (CRG) instrument operated by the CEA-Grenoble at the Institute Laue-Langevin (ILL, Grenoble), and using the 40-T cryomagnet developed by the LNCMI-Toulouse, the CEA-Grenoble and the ILL~\cite{40T_art}. Magnetic field pulses of up to 40~T, with a rise time of 23 ms and a total duration of about 100 ms can be produced every 10 minutes within this magnet. A crystal of 0.4~g was cut to fit the sample environment of the magnet of 7*6*7 mm$^3$ and mounted on a sapphire holder with $(h, h, l)_h$ in the scattering plane. Initial wavevector $k_{i}$ was fixed to 5.5~${\AA}^{-1}$. An analyzer was used to cut the background from the inelastic signal~\cite{in22}. Measurements were carried out at the base temperature of 2~K, in fields up to 39~T applied horizontally, with a tilt of 5$\degree$ from the $c$-axis of the crystal. 
Neutron counts were measured with a $^3$He fast single detector and simultaneously recorded by a digital data recorder that measures the voltage and current of the generator, as well as the voltage at the pick-up coil, allowing to synchronously store the time structures of the field signal and neutron count during a magnetic field pulse. Measurements were performed for some selected Bragg peaks by accumulating the data over several field pulses. After correction for the neutron time of flight, the field dependence of the intensities at each Bragg position has then been extracted by summing up the data, with either constant time or constant field-integration windows.

\section{\label{sec:two}Results and discussion}

\begin{figure}[hbt!]
\centering
\includegraphics[width=0.47\textwidth]{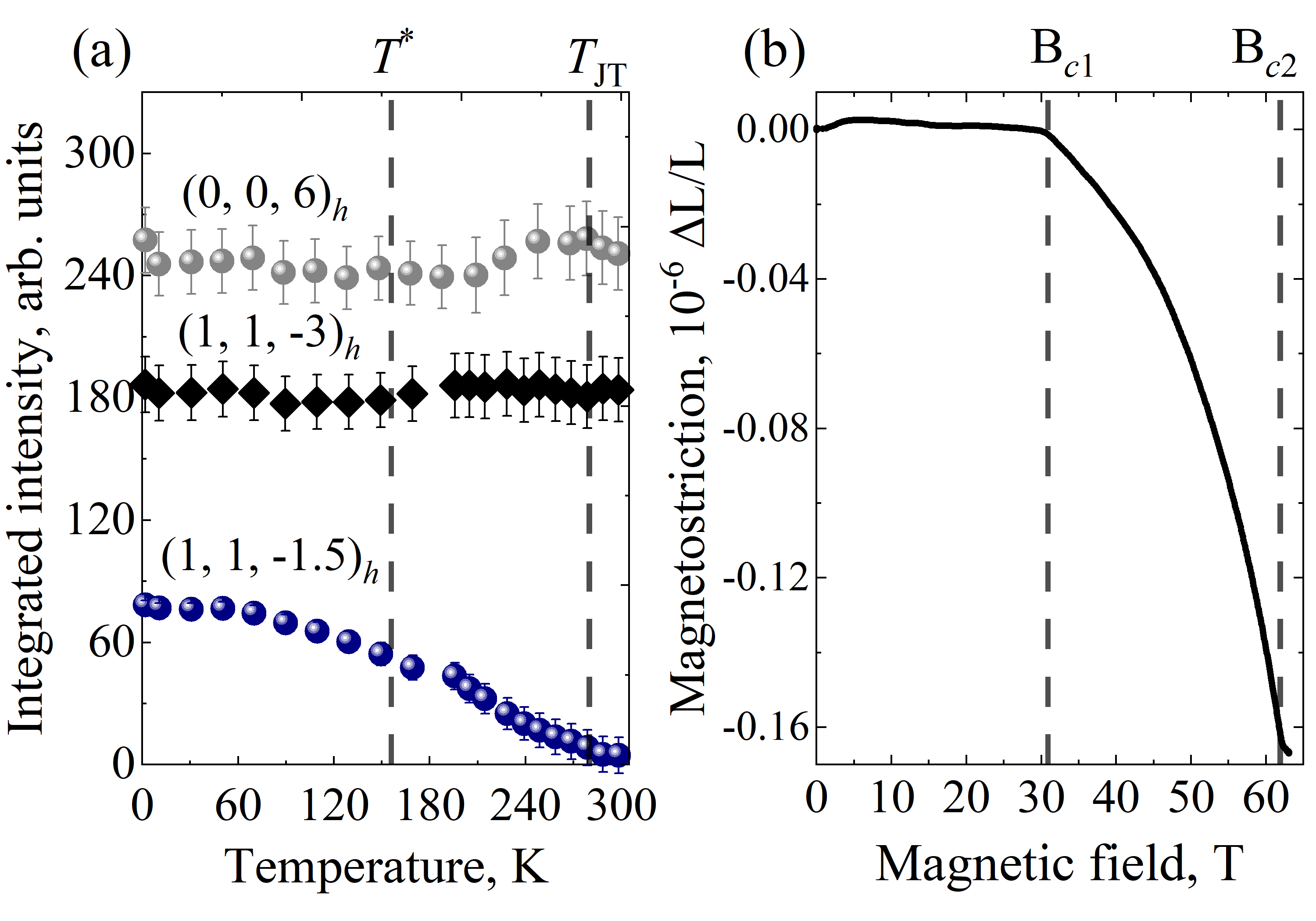}
\caption{(a) Temperature dependence of the integrated intensity of the (1, 1, -3)$_{h}$, (1, 1, -1.5)$_{h}$ and (0, 0, 6)$_{h}$ reflections. Dotted lines show both $T_{\text{JT}} = 285$~K and $T^{*} = 156$~K~\cite{dirk}. (b) Pulsed field magnetostriction data taken at 0.69~K, the magnetic field was applied along the $a$-hexagonal axis. Dotted lines indicate the critical magnetic fields, B$_{c1}= 30.9(4)$~T and B$_{c2}= 61.9(3)$~T.
}
\label{E4_Sr}
\end{figure}

Sr$_{3}$Cr$_{2}$O$_{8}$ is a Jahn-Teller active system with a structural transition from the rhombohedral $R\overline{3}m$ to the monoclinic $C2/c$ space groups. The Jahn-Teller transition occurs at $T_{\text{JT}} = 285$~K~\cite{chapon, wang_orb, dirk}.
The monoclinic distortion results in three twins, each rotated by $60\degree$ to each other around the shared hexagonal axis $c^{*}_{h}$. The transformation from hexagonal to monoclinic notation for these three reflections is given in Ref.~\cite{recalculation}. Further details on the structure can be found in Ref.~\cite{diana_mag}.
The formation of three monoclinic twins has been seen as a function of temperature by measuring key reflections in the (h, h, l)$_{h}$ plane (the subscript "$h$" denotes hexagonal notation), which correspond to a unique, double or triple twin. The measurements were done at three reflections: (1, 1, -3)$_{h}$, (1, 1, -1.5)$_{h}$ and (0, 0, 6)$_{h}$, where the (1, 1, -1.5)$_{h}$ peak is a new reflection corresponding to the monoclinic space group  [Fig.~\ref{E4_Sr}~(a)]. These results indicate that the structural transition develops gradually over the extended temperature range. The largest changes of the integrated intensities are observed for temperatures not only far below the Jahn-Teller transition temperature but far below characteristic temperature $T^{*} = 156$~K, that mark significant changes in the Raman spectra and is linked to a gap between the orbital excitations (see Ref.~\cite{dirk}).  Both ESR and Raman spectroscopy results conclude that strong lattice and orbital fluctuations are brought about by the Jahn-Teller distortion and compete with the system’s new structural phase down to temperatures well below the structural transition. Our results show that the monoclinic structural Bragg reflections only reach full intensity below approximately $T= 60$~K, the equivalent temperature of the intra-dimer interaction $J_0=5.5$~meV$ = 64.38$~K according to Ref.~\cite{diana_mag}. Due to the complexity of the signal and crystal quality, proper refinement of the low-temperature structure has not been reported. However, it is possible to analyze the physical properties of the compound by keeping the hexagonal notation, while taking into account any scattering signal due to monoclinic Bragg peaks as we do below. 

Sr$_{3}$Cr$_{2}$O$_{8}$ is a dimer system that remains in the paramagnetic state down to dilution-fridge temperatures. The ground state is a singlet (non-magnetic). The excited state is a gapped triplet that disperses along all directions in reciprocal space due to three-dimensional interdimer interactions. The zero-field behaviour of Sr$_{3}$Cr$_{2}$O$_{8}$ has been characterized in detail in Ref.~\cite{diana_mag}.  Magnetostriction measurements shown in Fig.~\ref{E4_Sr} (b), reveal three distinct phases. The first region, below the critical field B$_{c1} = 30.9(4)$~T shows a plateau in magnetostriction and can be associated with the paramagnetic phase. In the second region, between B$_{c1}$ and B$_{c2} = 61.9(3)$~T, the magnetostriction decreases rapidly and smoothly. The negative magnetostriction values reflect the sample contraction until the critical value B$_{c2}$ is reached, above which the sample length remains constant~\cite{Zapf_magnstr}. This phase is typically associated with an XY-AFM state \cite{gimr}, where spins spontaneously choose a particular orientation in the XY plane. The criticality at this point corresponds to that of Bose-Einstein condensation of magnons~\cite{aczel_sr}. Lastly, the magnetostriction plateaus above B$_{c2}$, is associated with a fully polarized paramagnet.  
After decreasing the field to 0~T, the lattice constants recover their original value. The overall magnitude of the magnetostriction is $\Delta{L}/L\approx10^{-6}$. This small value suggests that there is no significant restoration of the critical temperature to compensate for quantum fluctuations. These results are in agreement with those reported for continuous fields in Ref.~\cite{nomura2020}. The crystallographic axes shrink due to antiferromagnetic correlations mediated by Cr-O-Cr superexchange in the ordered phase.  Having understood the macroscopic details of Sr$_{3}$Cr$_{2}$O$_{8}$ phase diagram, microscopic details of the field-induced phases are discussed below.

Triplons or magnons can be regarded as bosonic pseudo-particles with $S=1$ and their degeneracy can be lifted by an external magnetic field. The external field can be modelled as a Zeeman energy that leads to the closing of the spin gap $\Delta$, populating the ground state. The dispersion relation of the magnon band  is defined as: 
\begin{equation}
\varepsilon_\mathbf{Q} \cong \sqrt{J_0^2 + J_0 \gamma(Q) } - g\mu_{B} B S_{z},
\label{disp_rel}
\end{equation}

where $\mathbf{Q}$ is wavevector, $J_0$ is the intradimer coupling and $\gamma(Q)$ is the Fourier sum of the interdimer interactions as described in Ref.~\cite{diana_mag}. The last term is the Zeeman term~\cite{gimr}, which controls the density of magnons. As the magnetic field increases, the excitation energy of magnons with S$^{z}=1$ is lowered, and eventually, it will cross zero energy. This approach applies only to fields below the critical field B$_{c1}$. 

\begin{figure}[hbt!]
\includegraphics[width=0.5\textwidth]{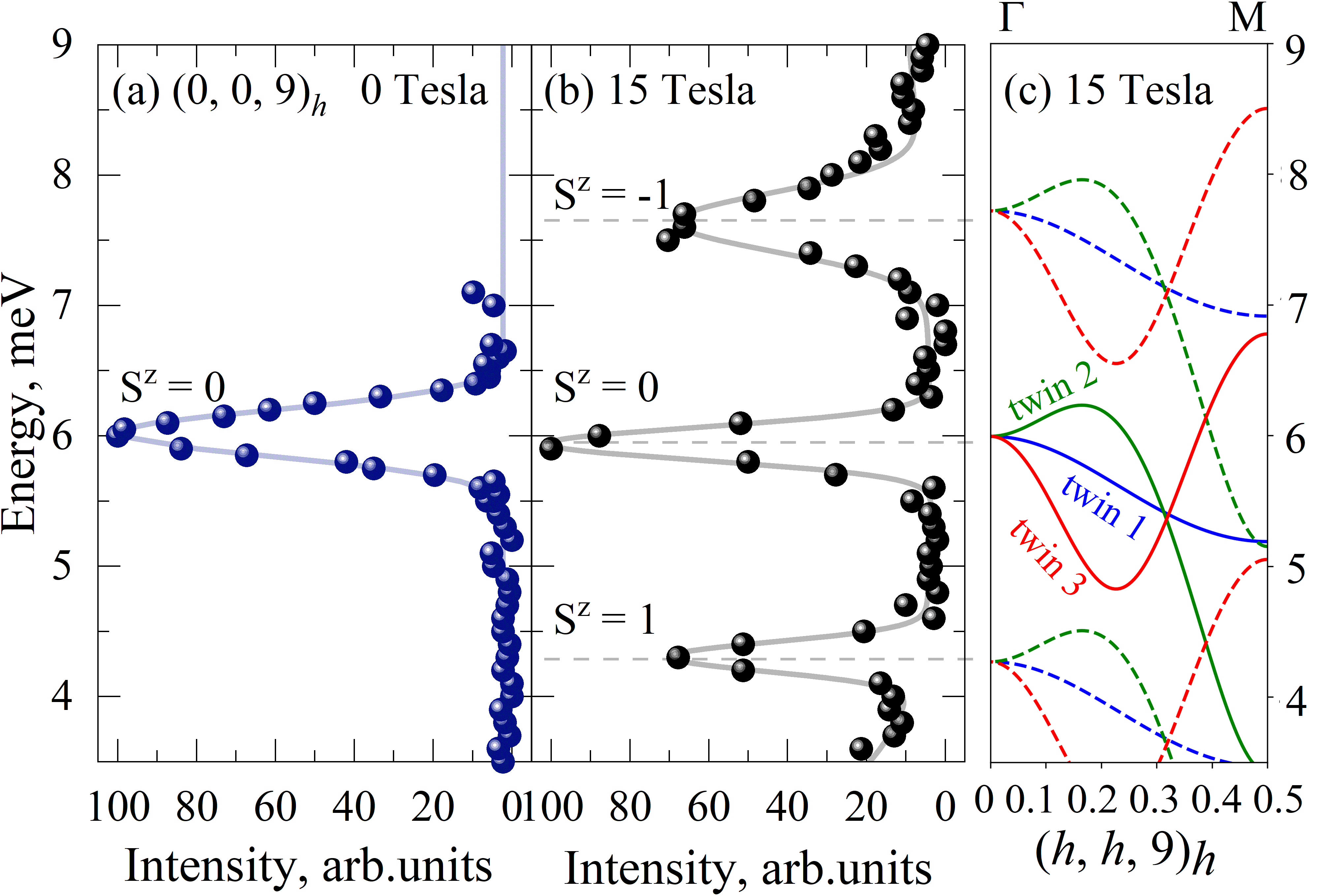}
\centering
\caption{The energy scan at the $\Gamma$-point (0, 0, 9)$_{h}$ at 2~K reveals (a) one peak in zero magnetic field and (b) three peaks at 15~T with the magnetic field along the $c$-axis.  (c) RPA model plus Zeeman energy (Eq.~\ref{disp_rel}) of the dispersion relation for the three crystallographic twins under an applied magnetic field of 15~T along the [$h$, $h$, 9]$_{h}$ direction. Dispersion relation and magnetic exchange interactions were extracted from Ref.~\cite{diana_mag}. The solid lines of the dispersion relate to S$^{z}=0$ which is independent of the magnetic field. The dispersion of the three twins merges at the $\Gamma$-point.}
\label{disp_scan_15T}
\end{figure} 

The microscopic details of the magnetic phase diagram of this compound have been investigated using neutron scattering and are compared to previously published bulk  properties~\cite{wang_esr,nomura2020}. Neutron energy loss scans at the $\Gamma$-point (0,~0,~9)$_{h}$ in zero magnetic field and 15~T applied along the $c$-axis, are presented in Fig.~\ref{disp_scan_15T}~(a, b). The triple degenerate magnon excitation at 6\,meV splits due to the applied field into three components S$^{z}=-1$, S$^{z}=0$ and S$^{z}=1$.  The small shift between the single peak at zero field and the middle peak (S$^{z}=0$) at 15~T correspond to a slightly different instrument configuration during these two measurements. 

Figure~\ref{disp_scan_15T}~(c) shows the dispersion relation, obtained by a random phase approximation (RPA) using the exchange interaction constants taken from Ref.~\cite{diana_mag} and by the Zeeman term as described in the equation above~\cite{gimr}.  The three coloured branches arise from the different coexisting structural domains as discussed above and shown in Ref.~\cite{recalculation}. The broadband of the excitations indicates relevant interdimer magnetic exchange interactions, as depicted in Fig.~\ref{disp_scan_15T}~(c). The magnon energies at the $\Gamma$-point for both fields are extracted using a Gaussian fit. Those results are used below in the construction of the magnetic phase diagram (Fig.~\ref{str}).

\begin{figure}[hbt!]
\includegraphics[width=0.45\textwidth]{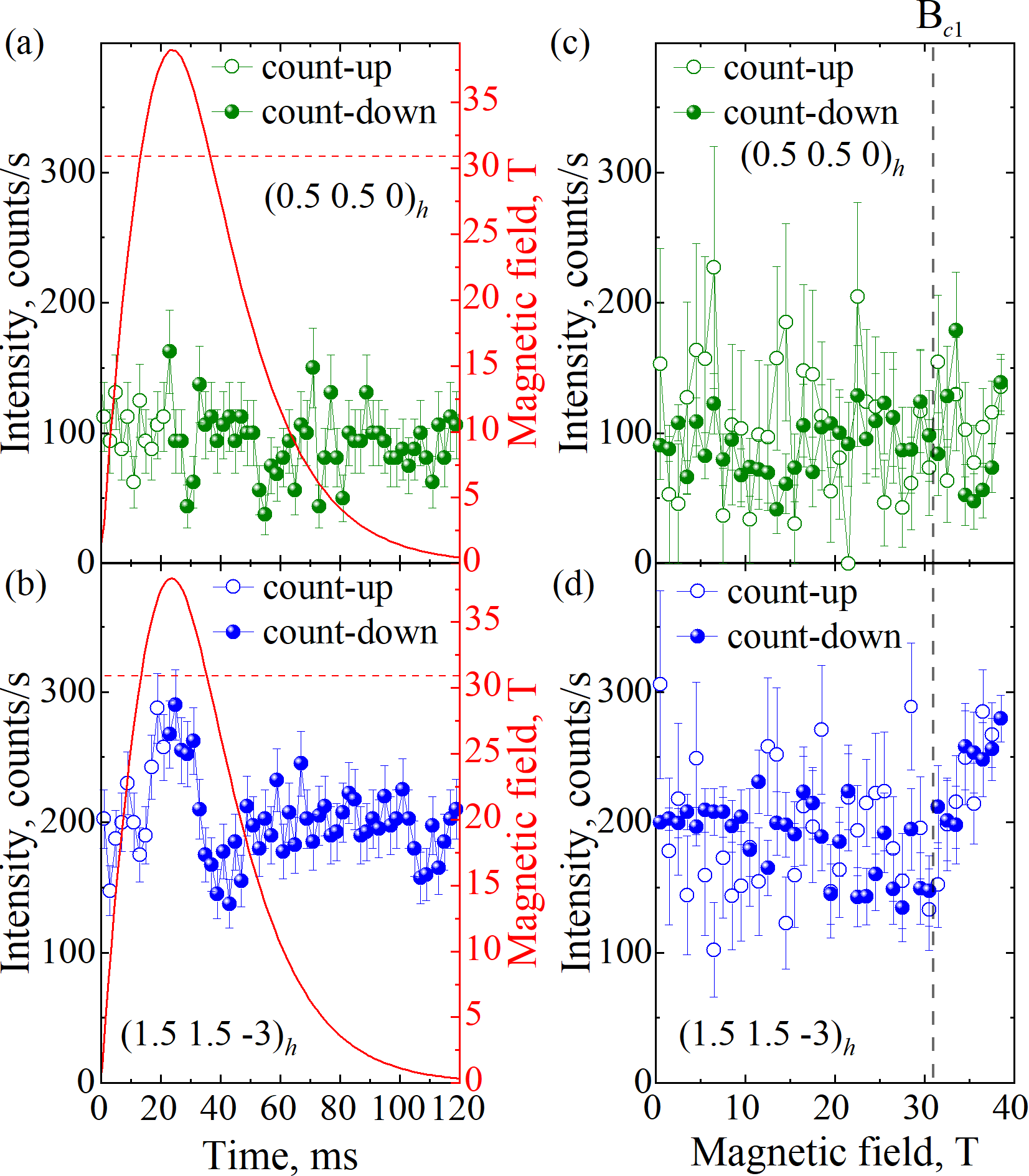}
\caption{
Time profile of a magnetic field pulse up to 38~T (red line) and time dependence measured at $T = 2$~K of the diffracted intensity at (a) (0.5, 0.5, 0)$_{h}$ and (b) (1.5, 1.5, -3)$_{h}$, integrated within time steps $\Delta{t} = 2$~ms. The right y-axis refers to the magnetic field values. Field dependence of the diffracted intensity at (c) (0.5, 0.5, 0)$_{h}$ and (d) (1.5, 1.5, -3)$_{h}$, integrated within field steps $\Delta{\text{B}} = 1$~T. The intensity is shown as the count-up and count-down points which are related to the rising and falling of the field pulse. In order to accumulate sufficient statistics, 80 and 220 pulses were done at (0.5, 0.5 0)$_{h}$ and (1.5, 1.5, -3)$_{h}$, respectively. The error bars $\Delta{I}$ are given by the square root of the neutron counts ($\Delta{I} = \sqrt{I/\tau}$, where $\tau$ is the total accumulation time over several pulses). The magnetic field has been applied along the $c$-axis with a tilt of 5\degree in order to reach the reflections.}
\label{pulses}
\end{figure} 

The minima of the magnon dispersion are neither located at the $\Gamma$-point nor at the border of the Brillouin zone, thus they cannot be probed by optical methods in pulsed fields, such as ESR. In the XY-AFM ordered state (equivalent to BEC), the bosons forming the condensate are represented by a coherent superposition between a singlet and the lowest $S_z = 1$ triplet states. In order to study the magnetic structure in this phase in Sr$_{3}$Cr$_{2}$O$_{8}$ we have performed neutron diffraction measurements in pulsed fields above B$_{c1}$ at the wavevectors corresponding to the minima of the triplon dispersion (M-point), namely (0.5, 0.5, 0)$_{h}$ and (1.5, 1.5, -3)$_{h}$ at 2~K. These results are presented in Fig.~\ref{pulses}. The time dependence of the magnetic field pulse is shown by the red-solid lines in the figure, from which the magnetic field distribution can be extracted. These results show zero magnetic intensity at (0.5, 0.5, 0)$_{h}$ and magnetic intensity at (1.5, 1.5, -3)$_{h}$ for fields above $\approx 31$\,T [see Fig.~\ref{pulses}~(b, d)]. 
Further measurements at other wavevectors were not possible to perform due to the restricted geometry of the magnet and the scattering conditions for wavevector transfer. Additionally, the expected magnetic reflections induced by the magnetic field in the XY-AFM phase coincide with monoclinic reflections, which complicates their separation. The observed magnetic signal confirms the field-induced phase transition starting at B$_{c1}= 30.9$\,T. The measured intensities are in agreement with the intensities of the measured and calculated RPA inelastic signal at those corresponding wavevectors~\cite{diana_mag}. In the case of the inelastic signal, the intensity is modulated by the dimer structure factor.

To understand this magnetic structure based on the information of the two measured peaks (one with zero intensity), representation analysis was performed using the BasIreps program implemented in the Fullprof suite~\cite{FP}. For these calculations, we have kept the high-temperature structure $R\overline{3}m$ as it has been done with the analysis of the inelastic data. This is a good approximation since together the three monoclinic twins keep the hexagonal symmetry. Small changes in magnetic exchange interactions and magnetic structures due to slight changes in interatomic distances are average out.  Additionally, there is a 5$^{\circ}$ tilt between the $c$-axis and the field direction. However, we do not consider a big impact of this tilt in our results due to the sample's broad mosaicity and the isotropic nature of the spin with reported g-factors $g_c=1.938(6)$ and $g_{ab}=1.950(1)$~\cite{aczel_sr}.

In the hexagonal structure, there is one inequivalent Cr$^{5+}$ site lying at Wyckoff position 6c, which generates two coupled Cr$^{5+}$ ions with antiparallel magnetic moments at ($0$, $0$, $z$).  The Bragg reflections indicating a magnetic structure can be described with the single propagation vector $k=$~(0.5,~0.5,~0).  
The magnetic representation at the Cr$^{5+}$ site with Wyckoff position along with the space group $R\overline{3}m$ and the propagation vector $k$ can be decomposed into a direct sum of irreducible representations as $\Gamma_{mag}(6c) = \Gamma_1 + 2 \Gamma_2 + 2 \Gamma_3 + \Gamma_4$. These representations contain 1, 2, $\overline{1}$ and {\it{m}} symmetry elements that leave $k=$ (0.5, 0.5, 0) invariant. Two of these solutions, $\Gamma_1$ and $\Gamma_3$, give a FM alignment between the dimers, which is only expected in the polarized phase above B$_{c2}$. Additionally, $\Gamma_3$ produces a strong Bragg peak at (0.5, 0.5, 0)$_{h}$, and $\Gamma_1$ produces an extremely weak peak at (1.5, 1.5, -3)$_{h}$ opposite to our observations.  

The irreducible representations, which allow an AFM coupling between the dimers with and without canting, are $\Gamma_2$ and $\Gamma_4$. The first one, $\Gamma_2$, has components on the $ab$ plane and along $c$-axis.  Meanwhile, the $\Gamma_4$ represents a structure with the moments on the $ab$ plane with zero component along the $c$-axis.  According to theory, in this phase, the spins are initially (at B$_{c1}$) on the plane perpendicular to the applied magnetic field ($c$-axis) with an angle of 180$\degree$ with respect to each other, and begin to tilt towards the field direction by an angle which depends on the strength of the external magnetic field and thus of the ground state population~\cite{Giamarchi1999, Tachiki1970_JPSJ, Tachiki1970_SPTP}.

\begin{figure}[hbt!]
\includegraphics[width=0.47\textwidth]{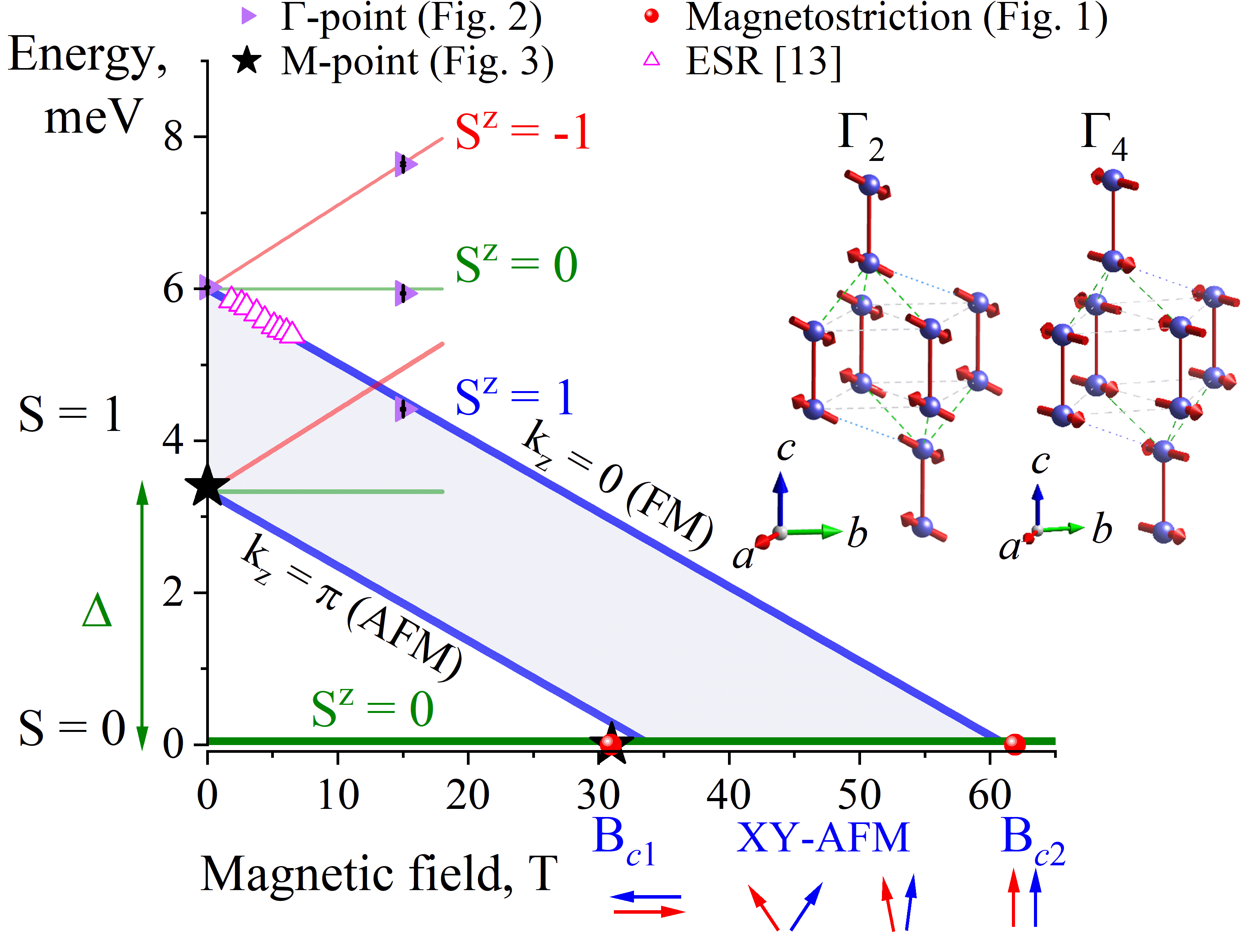}
\caption{
Zeeman splitting of the triplet modes in Sr$_{3}$Cr$_{2}$O$_{8}$. 
The $\Gamma$-point data are taken from the inelastic neutron scattering experiments (Fig.~\ref{disp_scan_15T}). The M-points data are from zero field inelastic neutron scattering and pulsed fields neutron diffraction measurements (Fig.~\ref{pulses}). All data were measured at $\approx2$\,K. ESR data points are taken from Ref.~\cite{wang_esr}. Critical fields are extracted from the magnetostriction results (Fig.~\ref{E4_Sr}). The red and blue arrows below the $x$-axis show a schematic illustration of the direction of the spins in a dimer.  
On the inset, the magnetic structure in the XY-AFM phase $\Gamma_2$ and  $\Gamma_4$ are presented~\cite{vesta}. The magnetic unit cell is 2$a \times $2$b \times c$, but only $a\times b\times c$ is shown for clarity. The red-solid line represents the dominant AFM intradimer interaction $J_0$. The interdimer interactions are shown by dotted lines.}
\label{str}
\end{figure}

The irreducible representation $\Gamma_4$ has two basis vectors: $S_{k_1}=(u, u, 0)$ and $S_{k_2}=(-u, -u, 0)$, where $u$ is pointing along the crystallographic $a$ and $b$ axis direction, and  zero component along the $c$-axis. In the irreducible representation $\Gamma_4$, the (1.5, 1.5, -3)$_{h}$ reflection is present while the (0.5, 0.5, 0)$_{h}$ reflection is absent, which is in good agreement with our measurements. This is also the case for $\Gamma_2$, which has two basis vectors: $S_{k_1}=(u, -u, v)$ and $S_{k_2}=(-u, u, -v)$, where $u$ is pointing along $a$ and $b$, and  $v$ is pointing along $c$. To distinguish between the two models, a comparison measurement needs to be done both at (0.5, 0.5, 3)$_{h}$ and (1.5, 1.5, -3)$_{h}$, as the ratios between the intensities of these two peaks are different for the two models. However, clean results from the measurements at  (0.5, 0.5, 3)$_{h}$ were not achievable due to the strong structural monoclinic peak which coincides with this wavevector. Therefore, with this limited information, we can conclude that both the canted AFM structure described by $\Gamma_2$ and the AFM structure describe by $\Gamma_4$ can be good representations of the XY-AFM phase at B$_{c1}$ where the FM, field polarized, component along the field direction is zero as depicted in the magnetic unit cells shown in the insert in Fig.~\ref{str} and below the x-axis in the same figure.

\section{\label{sec:conc}Summary}
Magnetostriction, inelastic neutron scattering and neutron diffraction have been used to establish the magnetic phase diagram of the spin-dimerized quantum antiferromagnet  Sr$_{3}$Cr$_{2}$O$_{8}$, shown in Fig.~\ref{str}.  Inelastic neutron scattering measurements in static magnetic fields up to 15~T reveals linear field dependency of the spin singlet-triplet excitations in this material, in agreement with the reported results of electron spin resonance measurements~\cite{wang_esr}.  A direct probe of the field-induced magnon condensate is achieved by doing neutron scattering measurements in a pulsed magnetic field up to 39~T. A representation analysis of the obtained data allowed us to derive two possible spin structures at 2~K above B$_{c1}$. Both of which confirm an antiferromagnetic configuration of the XY components that are perpendicular to the field direction. However, we cannot address on the ordering of the longitudinal spin components due to the limited reciprocal space available in our experimental techniques. The resolved XY-AFM configuration of the magnon condensation phase is consistent with the theoretical description, and agree with the reported results in other spin-dimerized compounds, such as TlCuCl$_{3}$~\cite{Tanaka2001} and the related isostructural compound Ba$_{3}$Cr$_{2}$O$_{8}$~\cite{kofu}.
Additionally, our zero-field diffraction results indicate that the monoclinic structure is only fully developed below 60\,K, the same temperature range of the intra-dimer interaction. This suggests that not only do the orbital and lattice degrees of freedom play a role in the stabilization of the crystallographic symmetry breaking transition, but that dimerization also has an important role. 
Similar behaviour has been reported for CuIr$_2$S$_4$~\cite{Bozin_Nat},  MgTi$_2$O$_4$~\cite{Yang_PRB} and NaTiSi$_2$O$_4$~\cite{Koch_PRL}.

\begin{acknowledgments} 
This work is based on experiments performed at the ILL: proposals 4-01-1493, 5-41-900 and 5-41-991. The latter measurements are available in Ref.~\cite{doi_541991}. Measurements were carried out at the E4 instrument at Helmholtz-Zentrum Berlin, beamtime 17106252-IN. We acknowledge support from the Deutsche Forschungsgemeinschaft (DFG) through SFB 1143 (Project No. 247310070), as well as the support of the LNCMI-CNRS and HLD-HZDR, members of the European Magnetic Field Laboratory (EMFL). A.G. has been supported by the Schweizerische Nationalfonds zur  F{\"o}rderung der Wissenschaftlichen Forschung (Grant No. 20-17555).
\end{acknowledgments}


\begin{thebibliography}{}
\bibitem{recalculation}
The reflection (1,1,-1.5)$_{h}$ corresponds to two monoclinic reflections (1,1,-1.5)$_{h}$ = (0,2,1)$_{m^{(1, 2)}}$ = (3,1,2)$_{m^{(3)}}$ ($m^{(n)}$ notation corresponds to $n^{th}$ monoclinic twin). Reflection (0,0,6)$_{h}$ corresponds to all monoclinic twins (0,0,6)$_{h}$ = (0,0,4)$_{m^{(1, 2, 3)}}$ and it is also an allowed hexagonal reflection, and the (1,1,-3)$_{h}$ reflection is allowed in both hexagonal and monoclinic and corresponds to the reflections (1,1,-3)$_{h}$ = (0,2,2)$_{m^{(1)}}$ = (3,1,3)$_{m^{(2, 3)}}$. Conversions from hexagonal to monoclinic Miller indices have been done using Ref.~\cite{diana_mag}.
\end{thebibliography}
\end{document}